# Title

MOSAIK: A hash-based algorithm for accurate next-generation sequencing read mapping

# Authors

Wan-Ping Lee[1], Michael Stromberg[1,2], Alistair Ward[1], Chip Stewart[1,3], Erik Garrison[1], Gabor T. Marth[1]
[1] Department of Biology, Boston College, Chestnut Hill, MA
[2] Illumina, Inc., San Diego, CA
[3] Broad Institute of Harvard and Massachusetts Institute of Technology, Cambridge, MA

# Abstract

MOSAIK is a stable, sensitive and open-source program for mapping second and third-generation sequencing reads to a reference genome. Uniquely among current mapping tools, MOSAIK can align reads generated by all the major sequencing technologies, including Illumina, Applied Biosystems SOLiD, Roche 454, Ion Torrent and Pacific BioSciences SMRT. Indeed, MOSAIK was the only aligner to provide consistent mappings for all the generated data (sequencing technologies, low-coverage and exome) in the 1000 Genomes Project. To provide highly accurate alignments, MOSAIK employs a hash clustering strategy coupled with the Smith-Waterman algorithm. This method is well-suited to capture mismatches as well as short insertions and deletions. To support the growing interest in larger structural variant (SV) discovery, MOSAIK provides explicit support for handling known-sequence SVs, e.g. mobile element insertions (MEIs) as well as generating outputs tailored to aid in SV discovery. All variant discovery benefits from an accurate description of the read placement confidence. To this end, MOSAIK uses a neural-net based training scheme to provide well-calibrated mapping quality scores, demonstrated by a correlation coefficient between MOSAIK assigned and actual mapping qualities greater than 0.98. In order to ensure that studies of any genome are supported, a training pipeline is provided to ensure optimal mapping quality scores for the genome under investigation. MOSAIK is multi-threaded, open source, and incorporated into our command and pipeline launcher system GKNO (http://gkno.me).

# Introduction

The widespread availability of next-generation sequencing platforms has revolutionized the life sciences through the ever more accessible ultra-high throughput DNA sequencing efforts (Drmanac et al. 2010). Next-generation sequencing technologies including Illumina, Complete Genomics, and Applied Biosystems (AB) SOLiD have been driving the current market forward, whereas Pacific Biosciences

SMRT (Eid et al. 2009), and Ion Torrent (Rothberg et al. 2011) are leading the development of third-generation sequencing instruments. These technologies bring novel opportunities for many applications including genetic variant discovery, epigenomic variant discovery, RNA-Seq, and ChIP-Seq, but also provide complex computational challenges. The short reads generated by these technologies are generally aligned to a reference genome as an early step in many of the current analysis workflows and the alignment quality limits the accuracy of any downstream analysis. Large sequencing projects often use sequencing machines from multiple manufacturers for data generation and can also make use of legacy data. It is desirable that any researcher tasked with analyzing the available data need not learn the intricacies of multiple alignment software packages to utilize all of the available data. This is unnecessary, since, MOSAIK can, uniquely, accurately align sequencing data from all current and legacy platforms.

Current sequencing technologies typically generate on the order of hundreds of millions of short reads (of the order of a few hundred nucleotides or shorter) on a single run. In order to analyze all of these reads in a reasonable amount of computational time, the performance of reference-guided alignment programs is paramount. The memory footprint of these algorithms must also be well managed to allow their deployment beyond institutions with extremely expensive computational infrastructure. These goals must be met without compromising the accuracy of the resulting alignments. Most existing aligners utilize hashing algorithms or the Burrows-Wheeler transform (Burrows et al. 1994; Cox et al. 2012) to search exact matches (algorithms may be modified to allow few mismatches) as their first step to achieve high performance and optimize memory usage. Theoretically, hashing method outperforms BWT method for DNA database searching (Boytsov 2011). The hash-based aligners, Eland (AJ Cox, Illumina, San Diego), MAQ (Li et al. 2008), mrFAST/mrsFast (Alkan et al. 2009; Hach et al. 2010), SHRiMP (Rumble et al. 2009; David et al. t2011), and ZOOM (Lin et al. 2008; Zhang et al. 2010) hash the reads and fit these hashes to the reference genome, while MOM (Eaves and Gao 2009), MOSAIK, PASS (Campagna et al. 2009), ProbeMatch (Kim et al. 2009), SOAP (Ruiqiang Li et al. 2008), SRmapper (Gontarz et al. 2013), and STAMPY (Lunter and Goodson 2011) hash the reference genome and store this for comparison with the reads. Major Burrows-Wheeler transform (BWT) based aligners include BWA (Li and Durbin 2009), Bowtie (Langmead 2010; Langmead and Salzberg 2012), segemehl (Hoffmann et al. 2009), and SOAP2 (Ruiqiang Li et al. 2009). In general, BWT-based aligners are sensitive but include a slow query operation (each FM-index query is slower than a hash query (P. Ferragina and G. Manzini; Paolo Ferragina and Giovanni Manzini 2001)). In regions with genomic variation (e.g. those regions in which the investigator is usually most interested), maintaining good performance generally leads to lower sensitivity (Gontarz et al. 2013; Mahmud et al. 2012). In addition, the Burrows-Wheeler transform method is less flexible than hash based methods. For example, it is more difficult for the Burrows-Wheeler transform to consider ambiguities by using IUPAC (Tipton 1994) ambiguity codes representing, for example, known SNPs. The main drawback of hash-based aligners is that they usually consume more memory than BWT-based aligners; however, as high-memory machines become cheaper, this becomes less of a problem. Currently, MOSAIK can be operated in a low-memory mode that keeps the memory footprint small (~8 GB for the human genome), ensuring that even for lower memory machines, MOSAIK can still be used with confidence.

Here, we introduce a reference-guided aligner, MOSAIK, that is highly sensitive, stable and flexible, whose utility on a range of different sequencing technologies has been demonstrated in the context of the 1000 Genomes Project (1000 Genomes Project Consortium 2010; 1000 Genomes Project Consortium et al. 2012). In addition to MOSAIKs ability to map data from all major sequencing technologies, it has been developed to address many of the issues currently facing genome researchers. These developments are outlined here. The primary goal of any mapping software is to minimize alignment artefacts and increase alignment sensitivity and accuracy. To achieve this, MOSAIK uses a Smith-Waterman algorithm and is able to align reads using IUPAC ambiguity codes, ensuring that alignments against known *single-nucleotide polymorphisms* (SNPs) are not penalized. Using this method, MOSAIK achieves positive predictive values (PPVs) of 99.5% for all alignments and 100.0% for high confidence alignments (those with a mapping qualities larger than 20) in experiments on simulated data. In addition to providing the genomic coordinates of the read mapping, it also important to provide a measure of the confidence in this coordinate. For this purpose, MOSAIK uses a neural-network based training scheme to provide well-calibrated mapping quality scores. In our experiments, the correlation coefficient between the quality scores assigned by MOSAIK and the actual scores is 0.97. To ensure that studies of any genome are supported, MOSAIK provides a training pipeline to ensure optimal mapping quality scores for the genome under investigation. A major area of active investigation is the study of structural variation (SV). MOSAIK has been designed to aid and simplify the discovery of such variants. In particular, known-insertion sequences, for example, mobile element insertions (MEIs), can be included as part of the reference genome. This helps to minimize alignment artefacts, but MOSAIK also provides a host of valuable information to the user on the paired-end reads that map to one of these sequences. When requested, MOSAIK also outputs all possible mapping locations for every read in a separate BAM file. This is essential for determining the mappability of the genome under study. Finally, MOSAIK is implemented in C++ as a modular suite of programs that is dual licensed under the GNU General Public License and MIT License. It is multi-threaded, open source, and incorporated into our command and pipeline launcher system GKNO (http://gkno.me).

# Results

### Alignments from all sequencing technologies

All of the available sequencing technologies use different techniques for library preparation, paired-end read protocols, and DNA sequencing, resulting in a range of read lengths, fragment lengths, base quality assignments, as well as different error profiles. Additionally, not all technologies report their sequencing reads in the conventional basespace (strings of the A, G, C and T nucleotides) format. Notably, AB SOLiD uses a di-base encoding scheme known as colorspace and single-molecule sequencing technologies use dark bases (Harris et al. 2008) for bases not registered by the instrument. These facts mean that all of the currently available aligners are tailored for use on data from one, or a small number of the available technologies. MOSAIK is the only aligner that can be used in a consistent manner across all of these technologies.

In addition to the second-generation technologies, Illumina, Roche 454, and AB SOLiD, MOSAIK can also be deployed on third-generation technologies, in particular, Pacific Biosciences and Ion Torrent reads. MOSAIK uses the same algorithmic approach for all sequencing technologies, however, since the characteristics of each technology are different, the resultant alignment rates vary, as shown in Table 1.

*Lee et al.*

These alignment rates were generated using Illumina paired-end (PE), single-end (SE) and Roche 454 SE reads generated using the MASON read simulator (http://www.seqan.de/projects/mason/) as well as Illumina and AB SOLiD reads from the Han Chinese in Beijing (CHB) population from the 1000 Genomes Project. For the third-generation technologies, we used *E. coli* reads provided by Ion Torrent (Ion Torrent E. coli) and *V. cholerae* reads provided by Pacific Biosciences (Pacific Biosciences *V. cholerae*).

**Table 1:** Summary of the alignment accuracies achieved by MOSAIK for reads generated from different sequencing technologies. With the exception of the Pacific Biosciences data, all alignments were generated using MOSAIK's default parameters.

| Technologies | Aligned (%) | Speed (reads/second) | Read lengths [min;max] | Reference genome | Dataset |
| --- | --- | --- | --- | --- | --- |
| Illumina; PE | 99.98 | 83.95 | 100; 50 | Human hg19 | MASON simulated |
| Illumina; SE | 99.75 | 153.98 | 100; 76; 50 | Human hg19 | MASON simulated |
| Illumina; PE/SE | 91.48 | 147.42 | 81; 76; 51; 45; 41 | Human hg19 | CHB population in 1000G |
| 454; SE | 99.42 | 8.018 | 400.673 [266;529] | Human hg19 | MASON simulated |
| Ion Torrent | 77.02 | 20.85 | 223.99 [59;398] | *E. coli* strain 536 | Ion Torrent released |
| SOLiD | 55.64 | 126.81 | 50 | Human hg19 | CHB population in 1000G |
| Pacific Biosciences* | 85.79 | 0.69 | 698.61 [48;6084] | V. cholerae 4,033,464 bp. | Pacific Biosciences released |

*The parameter set "-hs 10 -mmp 0.5 -act 15" was used as opposed to the default values "-hs 15 -mmp 0.15 -act 55".

In general, sequencing reads containing fewer sequencing errors have higher alignment rates, e.g. Illumina reads, and longer or paired-end reads require more time to align. That paired-end reads take additional time is not unexpected. If one of the reads in a pair cannot be mapped unambiguously, additional searches are performed guided by the mapped mate in the pair. The additional processing time results in more accurate alignments as well as a lower fraction of unaligned reads. AB SOLiD reads are aligned in colorspace (converting to basespace prior to alignment loses all of the benefits of colorspace), but additional processing is required due to the required conversion of the alignments into basespace post-alignment. These experiments show that MOSAIK works well for existing sequencing technologies.

## Highly accurate alignments on simulated data

To investigate the accuracy of reads aligned using MOSAIK, we simulated a total of 120 million Illumina paired-end reads from chromosome 20 of the Hg19 human genome using the MASON read simulator. Reads of length 76 and 100 basepairs were simulated with a haplotype SNP rate of 0.1%. The reads were aligned against the entire human genome using BWA-0.5.9, BOWTIE-2.0-beta5, STAMPY-1.0.13, and MOSAIK-2.1.78. The default parameter settings were used for all of the aligners. The positive predictive value of each aligner was then calculated as the number of correctly placed reads (the genomic coordinate of the mapped read agreed with the known location of the read from MASON) divided by the total number of mapped reads. Notice that an alignment is considered incorrect as the aligned position is out the 20bp tolerant window and thus too short alignments may be considered as incorrect. We choose 20bp as the tolerant window since most of alignments contain fewer than 20bp clipped bases (see supplemental Figure S1).

Figure 1 shows the positive predictive value (PPV, the number of correctly mapped reads divided by the total number of mapped reads) of the aligners as a function of mapping quality cutoffs (complete information is shown in Figure S2). At a mapping quality cutoff of twenty, for example, the PPV is calculated using only those reads with mapping quality values greater than or equal to twenty. It can be seen that the PPVs of BWA and MOSAIK are comparable and are significantly better than those achieved by BOWTIE and STAMPY. For mapping quality cutoffs smaller than five, BWA is more accurate (fewer incorrect alignments among total mapped alignments) than MOSAIK, however, MOSAIK is the most accurate as the mapping quality cutoff is increased. For a mapping quality cutoff of twenty (a common cutoff employed by downstream analysis tools that only wish to consider confidently aligned reads), the PPVs of MOSAIK, BWA, BOWTIE and STAMPY are 100.00%, 99.99%, 99.79% and 99.63% respectively. These results are summarized in Table 2. While the accuracy achieved by MOSAIK on the simulated data is only marginally higher than for the other compared aligners, in regions with low complexity or genetic variants, this improvement could be significant. All the aligners have no problem aligning to regions of the genome with little variation, it is the accurate placement of reads in variant regions that are of most interest.

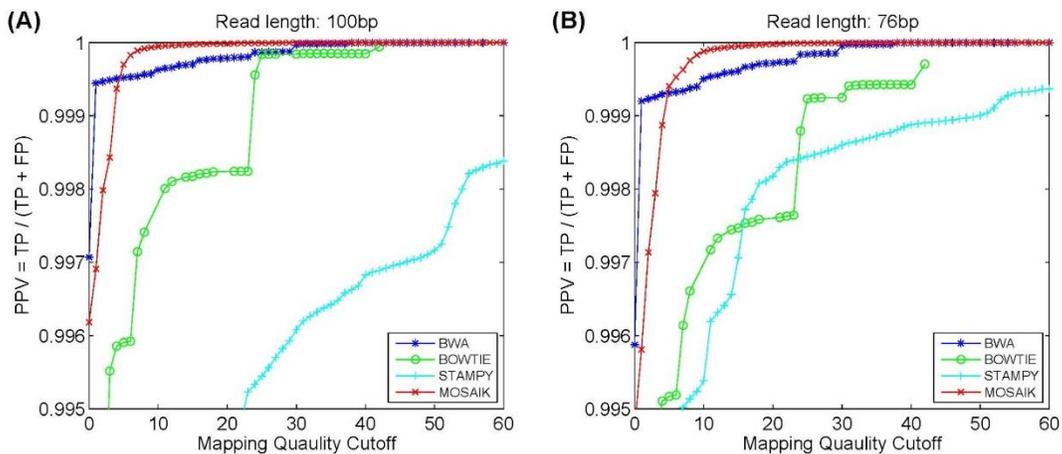

**Figure 1:** The positive predictive value of aligners (the number of correctly mapped reads divided by the total number of mapped reads) as a function of mapping quality threshold. Datasets in (A) 100bp and (B) 76bp read lengths. PPV, TP, and FP stand for positive predictive value, true positive, and false positive, respectively.

**Table 2:** The positive predictive values (PPV) in terms of mapping quality cutoffs

| MQ cutoffs | 30 | | 20 | | 10 | | 0 | |
|---|---|---|---|---|---|---|---|---|
| Read lengths | 100 | 76 | 100 | 76 | 100 | 76 | 100 | 76 |
| BWA | 1 | 1 | 0.9998 | 0.9997 | 0.9996 | 0.9995 | **0.9971** | **0.9959** |
| BOWTIE | 0.9998 | 0.9992 | 0.9982 | 0.9976 | 0.9980 | 0.9972 | 0.9823 | 0.9819 |
| STAMPY | 0.9961 | 0.9986 | 0.9945 | 0.9982 | 0.9897 | 0.9954 | 0.9813 | 0.9909 |
| MOSAIK | **1** | **1** | **1** | **1** | **0.9999** | **0.9999** | 0.9962 | 0.9947 |

Figure 2 shows receiver operating characteristic (ROC) curves for the same data. The total number of mapped reads (x axis) is plotted against the number of incorrectly mapped reads (y axis). Each point on the curve represents the number of alignments whose mapping qualities are greater than or equal to the indicated value. MOSAIK has a relatively smooth curve, ensuring that downstream tools that employ mapping quality cutoffs (i.e. ignoring all reads with mapping qualities less than the cutoff) do not incur extremely large changes in the number of reads while progressively increasing the cutoff. Conversely, the other aligners do not share this property. For example, consider the BWA alignments. By decreasing the mapping quality cutoff from 30 to 29, the number of incorrectly mapped reads increases by 308.56% while for MOSAIK, the increase is a much more modest 6.25%. Downstream analysis tools require a useful mapping quality scale, so that excluding lower quality reads improves the specificity of the analysis results. The dynamic range demonstrated by MOSAIK is therefore a very valuable result for these tools.

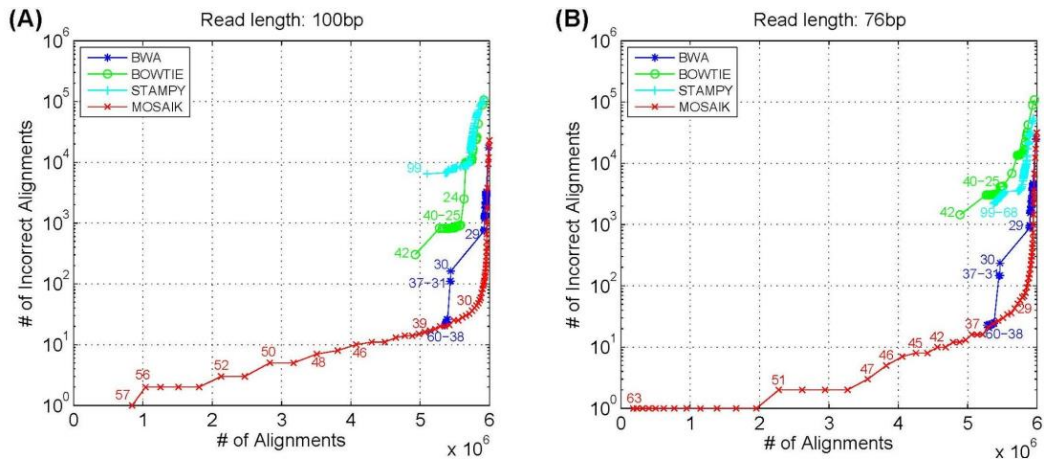

**Figure 2:** The receiver operating characteristic (ROC) curves; Datasets in (A) 100bp and (B) 76bp read lengths. Each point represents the total numbers of alignments whose mapping qualities are greater than the indicated value. MOSAIK has a relatively smooth curve, ensuring that downstream tools that employ mapping quality cutoffs (i.e. ignoring all reads with mapping qualities less than the cutoff) do not incur extremely large changes in the number of reads while progressively increasing the cutoff.

## Mapping quality calibration

The Phred mapping quality score present in the standard SAM/BAM format represents the probability that the read was mapped incorrectly and is defined as:

$Q = -10 \log_{10} P$     (Equation 1)

, where $Q$ is the Phred score and $P$ is the probability that the read was misaligned. For example, a read assigned a Phred mapping quality score of 30 has a 1 in 1000 chance of being misaligned.

MOSAIK's mapping qualities are obtained using a neural network that approximates the error function when provided with features such as best and second best Smith-Waterman alignment scores, read entropy, number of potential mapping locations and hashes. For paired-end reads, the fragment length for mapped paired end reads is also used in the neural net to produce more precise mapping quality calculations. MOSAIK embeds the Fast Artificial Neural Network (FANN) library (http://leenissen.dk/fann/wp/), which implements multilayer artificial neural networks in C, supporting both fully connected and sparsely connected networks, to calculate Phred score for each alignment.

The default neural network provided with MOSAIK was generated by training on the human genome. The first step involves simulating reads and then aligning them to the human reference genome to obtain MOSAIK's behaviour such as best and second best Smith-Waterman scores, entropies of reads, numbers of obtained mappings and hashes. Then, the neural network was trained based on MOSAIK's behaviour.

Figures 3(A) and 3(B) compare the actual (calculated using Equation (1)) and the assigned mapping quality scores. Both, BOWTIE and MOSAIK produce very accurate Phred score mapping qualities across the whole quality score spectrum. The Pearson correlation coefficients between the assigned and actual quality scores are shown in Table 3. MOSAIK has an average (across all read lengths investigated) correlation coefficient of 0.9698, compared with 0.9061, 0.9207, and 0.8652 for BWA, BOWTIE, and STAMPY respectively.

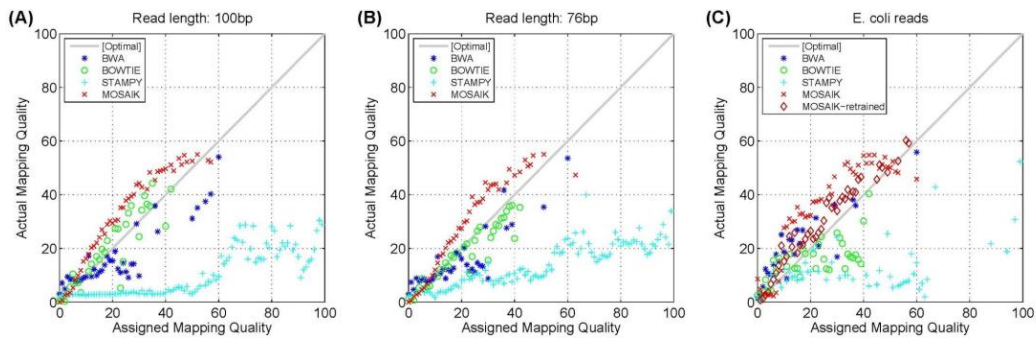

**Figure 3:** The correlations between the aligners' assigned and actual mapping qualities. Phred score scheme. (A) and (B) simulated datasets in 100bp and 76bp read lengths. (C) *E. coli* simulated dataset in which "MOSAIK" is MOSAIK's default mapping-quality network trained by human genome while "MOSAIK-retrained" is the retrained mapping-quality network by using *E. coli* simulation and *E. coli* genome. The detailed numbers of the Pearson's correlation coefficients are given in Table 3.

**Table 3:** Pearson's correlation coefficients of mapping qualities

| read length | 100 | 76 | *E. coli* |
|---|---|---|---|
| BWA | 0.8987 | 0.8625 | 0.8936 |
| BOWTIE | 0.9027 | 0.9449 | 0.6989 |
| STAMPY | 0.8317 | 0.8818 | 0.5262 |
| MOSAIK | **0.9609** | **0.9497** | 0.8881 |
| MOSAIK-retrained | -- | -- | **0.9749** |

## Retraining Mapping-Quality Neural Network for *E. coli* Alignment

The genomes of different species differ in many respects including sequence content (base composition as well as relative frequency of repeat or low-complexity sequence) as well as the size of the genome. Most aligners, including MOSAIK, are general programs that can operate on any given reference genome, however, in general, the properties of the genome under investigation are ignored. MOSAIK provides a retrainable mapping-quality pipeline to generate applicable neural networks for different genomes or sequencing technologies. This means that the calibration of the mapping quality scores remains of a very high quality, regardless of reference genomes.

To demonstrate the merit of the retrainable mapping-quality pipeline, we used 6 million simulated paired-end reads from the *E. coli* genome to train a neural network (see supplemental methods: Retraining Mapping Quality Neural Network). An additional independent set of 6 million simulated *E. coli* paired-end reads were then generated and aligned to the *E. coli* genome using multiple aligners. The assigned and actual mapping quality scores are plotted for all aligners in Figure 3(C). There are two sets of data for MOSAIK: the first (red crosses) is generated using the default neural network trained on the human genome, and the second (dark red diamond) uses the neural network retrained on the *E. coli* genome. It is clear that the mapping qualities generated by the retrained neural network for MOSAIK are the best calibrated, although the data using the human genome trained neural network is still of a high quality. Also of note, Figures 3(A) and 3(B) show that BOWTIE has quite well calibrated mapping qualities for mapping to the human genome, however, when applied to *E. coli*, the calibration is noticeably worse.

## MOSAIK accurately accounts for short INDELs

MOSAIK uses a Smith-Waterman (SW) algorithm as the final polishing step to produce pairwise read alignments, which is the preferred choice for aligning gapped (short INDELs) sequences since it seeks all possible frames of alignment with all possible gaps. To assess the sensitivity of different aligners to short-INDELs, we simulated Illumina paired-end reads containing 1-14bp INDEL events that are generated by a genome simulator, MUTATRIX (https://github.com/ekg/mutatrix). For each INDEL length, we introduced an average of 100 events, with approximately 800 spanning reads (see supplemental Figure S4).

Figures 4(A) and 4(B) plot the sensitivity (number of correctly mapped reads divided by the total number of simulated reads) as a function of the INDEL length. An alignment is considered correct when it is mapped to the correct position and contains the simulated variant. Alignments containing the correct variants can facilitate downstream variant detectors detecting variants depending on alignments and need no any realignment step which is timing consuming. MOSAIK is the most sensitive aligner considered here when considering deletions. When considering insertions, MOSAIK's sensitivity is comparable to, but slightly worse than those of STAMPY and BOWTIE. It is clear from Figures 4(A) and 4(B) that MOSAIK is the only mapper considered here that is highly sensitive to both insertion and deletion polymorphisms. We understand that some aligners tend to report partial alignments that may not contain variants but are mapped to right places. Those alignments still provide values for variant detections. We thus change the criteria of correct alignments used in Figures 4(A) and 4(B). In figures 4(C) and 4(D), an alignment is considered a correct mapping when it is entirely or partially mapped to the correct positions. The four aligners achieve 96% sensitivity based on the criteria.

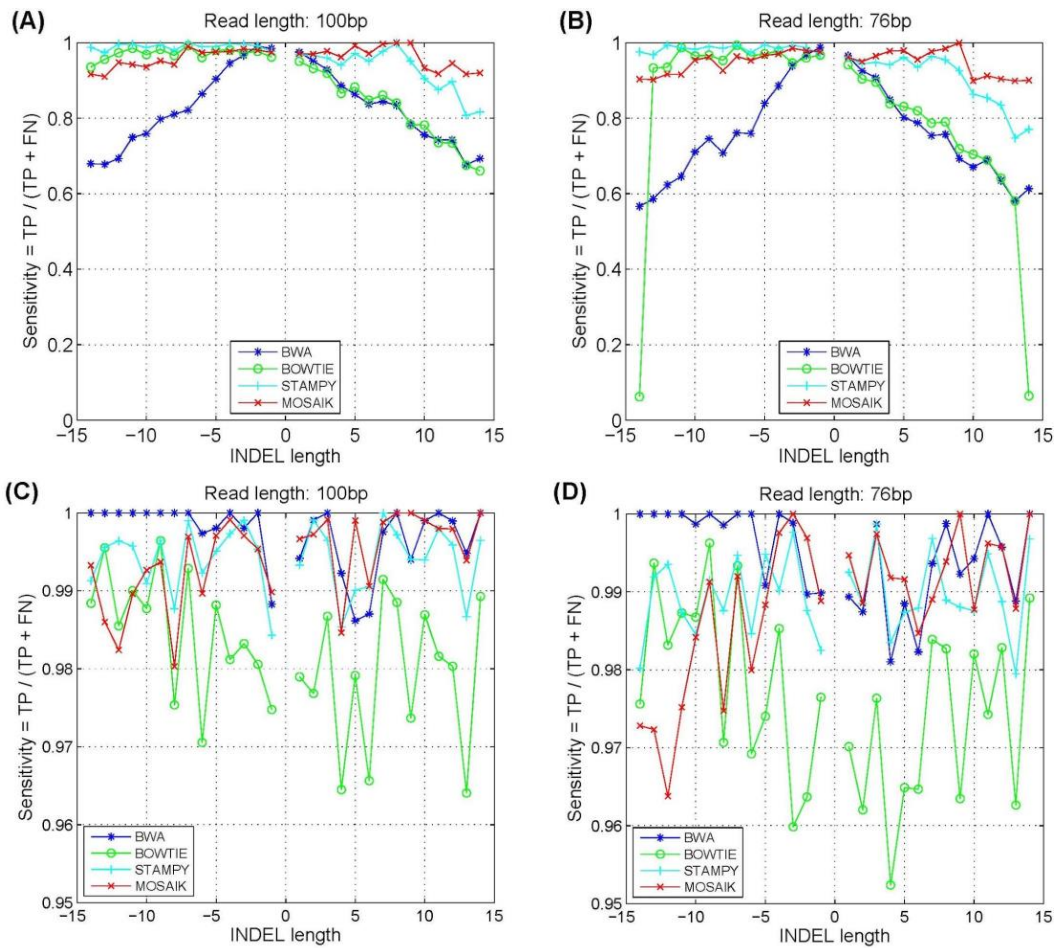

**Figure 4:** The sensitivities of simulated reads that cross INDELs, which is defined as the number of correct mapped reads divided by the number of simulated reads for each INDEL length. In (A) and (B), the alignments are considered correct as they cross INDELs, while in (C) and (D), the alignments are considered correct as they are entirely or partially mapped to the correct positions. TP and FN stand for "true positive" and "false negative".

## Effect of mapping errors on SNP studies

Aligners provide information on where reads map in the human genome along with information on the confidence of the mapping, however, they do not themselves weigh evidence for genetic variants in the genome being studied. Dedicated variant callers use the information provided by the mapper in statistical models to determine if there is enough evidence to report a difference with respect to the reference genome. To determine the effect of the mapping on single nucleotide polymorphism (SNP) discovery, we simulated 1,486 SNPs on the human chromosome 20 using MUTATRIX. We then used MASON to generate 12 million reads (with read lengths of 76 and 100 basepairs) from this mutated chromosome. The same four aligners were then used to align these reads back to the human reference genome and the variant callers FREEBAYES (Garrison and Marth 2012) and SAMTOOLS (Li 2011) were used to call SNPs. Figure 5 shows the variant callers sensitivity to SNPs as a function of the false discovery rate (FDR) (the complete information is shown in Figure S3). The points on the curves are generated by only considering SNP calls with variant quality scores (provided by the variant caller) greater than a specific cutoff. Moving from right to left, SNP calls with lower quality scores are cumulatively being included. The points closer to the upper-right corner have lower variant quality scores which contains more true positive as well as false positive. Both FREEBAYES and SAMTOOLS produce lower sensitivity calls on the BOWTIE alignments and have a lower FDR on BWA and MOSAIK alignments. It is clear from both Figure 5(A) and 5(B) that the most sensitive SNP calls are produced when using the MOSAIK alignments, although the BWA alignments are also of a high quality. It is also worth noting that the SNP calls produced by FREEBAYES are more sensitive than those produced by SAMTOOLS regardless of the mapper used.

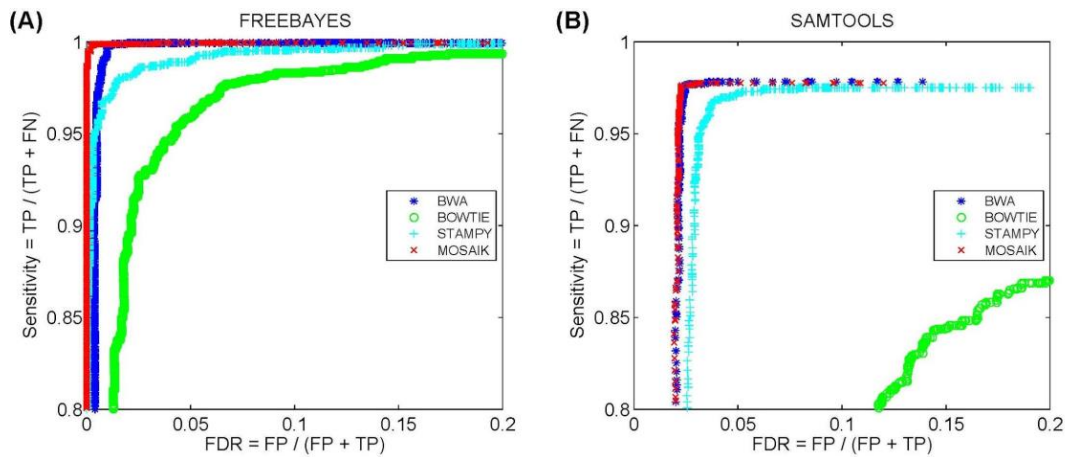

**Figure 5:** The receiver operating characteristic (ROC) curves of SNPs called by FREEBAYES and SAMTOOLS. The points on the curves are sorted by called qualities and the points closer to the upper-right corner have higher called qualities. The true positive (TP), false positive (FP), and false negative (FN) are calculated by intersecting SNPs called on each aligner's alignments and gold SNPs called on the simulated alignments.

## MOSAIK provides explicit support for structural variation detection

Detecting structural variations using NGS data is a more complex task than detection of short variants and often requires or would benefit from information over and above that ordinarily required for small variant detection. An increasing number of SV detection algorithms are being developed and, in order to increase the effectiveness of these algorithms, MOSAIK has been developed to provide as much relevant and useful information as possible.

There are many genetic sequences that can be considered distinct from the standard set of chromosomes in the genome under investigation. These can include repetitive sequences such as mobile elements (Prak and Kazazian 2000), viruses (e.g. human endogenous retroviruses (Griffiths 2001)), known novel insertions (Costantini and Bernardi 2009; Levy et al. 2007) or bacterial contaminants (Osoegawa et al. 2001) amongst others. MOSAIK provides support for an additional reference genome file containing any genetic sequences provided by the investigator. The advantages of this are two-fold: a) reads originating from contaminants will map to the additional sequences, rather than a lower quality mapping to the best location in the standard reference genome. These sequences essentially act as a sink to catch all the reads that do not originate from the standard reference, reducing the number of mismapped reads that the variant detectors have to contend with. b) Reads mapping to repetitive elements (e.g. ALU or LINE elements) are identified as mapping to the additional reference sequence. MOSAIK reports the coordinates of the best mapping in reference genome coordinates, but also includes an additional tag in the BAM file (appearing as ZA in the BAM file), indicating that the read maps to one of the additional reference sequences. Our MEI detector, TANGRAM (J Wu, W-P Lee, A Ward, E Garrison, M Konkel, M Batzer, G Marth, in preparation) looks for read pairs with one mate uniquely aligned to the genome and the other mate falls within a mobile element reference sequence. Only relying on this information provided by MOSAIK, the sensitivity of MEI detection can achieve 84%.

Additionally, this ZA tag provides valuable information to minimize work for any SV detection. Most SV detectors make extensive use of information from paired end reads. If the two mates in a pair map to greatly separated locations (often the case when the read pair spans or falls within a structural variant), multiple searches through the BAM files are required to assemble all of the information about both mates. This can be a lengthy task, severely impacting the performance of SV detectors. The ZA tag provides a host of information about the reads mate, including the location, mapping quality, number of mappings for the mate, which ensures that these searches are not required, created vast increases in the efficiency of the SV detectors using this information.

Many genomes contain regions that are considered unmappable, usually due to the presence of low complexity DNA. Depending on the algorithms employed, NGS reads can still map to these regions; however, it is often prudent to omit these reads from variant detection. Instead of discarding reads mapping to multiple locations, or picking the best quality alignment, MOSAIK records all locations to which a read maps (given the constraints imposed by the selected parameters) and records them in a separate BAM file. Since the number reads mapping to multiple locations as well as the number of entries for each multiply aligned read can be extremely large, the resulting BAM file has the potential to be excessively large. By default, MOSAIK omits much of the read specific information (e.g. read name, sequence and error information), allowing for effective compression of the file after positional sorting, resulting in very small BAM files. The information contained in these BAM files allows easy identification of genomic regions where many individual reads are aligning. These regions are those that can be considered unmappable, since reads hitting these regions are also able to align to other genomic regions. Thus they provide a guide to the mappable genome which can greatly aid in variant discovery.

# Applications

## SNP and INDEL Analyses in the 1000 Genomes Project

The 1000 Genomes Project is in the process of using second-generation sequencing instruments to study human genetic variation at the population level. The Phase I (1000 Genomes Project Consortium et al. 2012) release, based on a population of 1,092 sample individuals in 14 populations includes approximately 38 million SNPs, 1.4 million bi-allelic INDELs, and 14,000 large deletions. These calls were generated from approximately 966 billion reads and 64 trillion base pairs of human DNA and were sequenced using Illumina, AB SOLiD, and Roche 454 for both low-coverage whole-genome and exome targeted sequencing data. A collaborative effort between Boston College and the National Center for Biotechnology Information (NCBI) used MOSAIK to align all of the reads from all of these machines, and served as the official primary alignment set for the exome sequencing data (Marth et al. 2011) and an alternative alignment set for the low-coverage.

Based on the MOSAIK alignments, SNP, MNP (multi-nucleotide polymorphism) and INDEL calls were generated using the FREEBAYES Bayesian variant calling software. 33,324,407 SNPs were detected in the autosomes of the 1,092 samples, of which, only 23.8% were previously known sites (contained in dbSNP). The transition/transversion (ts/tv) ratio for these sites was 2.12 (2.1 for novel sites and 2.17 for known sites). The Illumina exome data yielded 344,781 SNPs with a ts/tv ratio of 3.18 (3.09 for the novel sites and 3.52 for the known sites) and 22.1% of the exome sites were previously known. The SOLiD exome data yielded 176,637 SNPs with a ts/tv ratio of 3.34 (3.22 for novel sites and 3.58 for known sites). The ts/tv ratios are in accordance with expectations for both the low-coverage and the exome SNPs.

## Other SNP Studies

In addition to the 1000 Genomes Project, MOSAIK is widely used for other human clinical genome studies, such as human cancer studies (Su et al. 2012; Kathryn G Roberts et al. 2012; Lin et al. 2012; Wang et al. 2011; Chung et al. 2011; Goya et al. 2010). MOSAIK is also used for other species genome studies including model species (Cridland and Thornton 2010; Hillier et al. 2008), HIV (Henn et al. 2012; Malboeuf et al. 2012; Campbell et al. 2011; Wilen et al. 2011), parasites (Farrell et al. 2012; Dark et al. 2012, 2011), plants (Iorizzo et al. 2011; Neves et al. 2011; Cannon et al. 2010), and other animals (Aslam et al. 2012; Fraser et al. 2011).

## Human mobile element insertion discovery

In addition to short variants, the 1000 Genomes Project aims to characterize larger structural variations present in the human population. By augmenting the reference genome with known mobile element insertions (MEI), the MOSAIK alignments were able to provide a host of information about their distribution in the human population. As part of the pilot phase of the project, 7,380 MEI polymorphisms were detected using the whole-genome sequencing data (Stewart et al. 2011). This sample set included 60 samples of European origin (CEU), 59 African (YRI), and 60 Asian samples from Japan and China (CHB/JPT). The FDRs for Alu, L1, and SVA insertions were 2%, 17%, and 27% respectively. In Phase I of the project, the co-submitted TANGRAM software package was used to call MEIs in the...

# Discussion

MOSAIK is a highly sensitive, stable and flexible reference-guided read mapper which supports all existing sequencing technologies, e.g. Illumina, Roche 454, AB SOLiD, Pacific Biosciences, and Ion Torrent. While MOSAIK is extremely accurate (positive predictive values achieve 99.5% for all alignments and 100.0% for alignments whose mapping qualities are larger than 20 on simulated data), not all reads are aligned with equal confidence. The mapping qualities that MOSAIK provides are generated using a retrainable neural network and are a very good representation of the probability of the alignment being incorrect. In fact, the correlation coefficient between MOSAIK assigned and the actual mapping qualities is 0.97. The retraining pipeline ensures optimized mapping quality score schemes for any genome being studied. For example, when considering aligning against the *E. coli* genome, the correlation coefficient increases from 0.89 to 0.97 when using the human and the *E. coli* neural nets respectively. By using the Smith-Waterman algorithm, MOSAIK is very effective at mapping reads containing short INDELs, and the experiments demonstrate that the sensitivity of INDEL mappings is greater than 90%. Additionally, MOSAIK provides explicit support for SV (in particular MEI) detection.

The default parameters used by MOSAIK were optimized using simulated Illumina datasets from the human genome. They were generated to provide a balance between mismatches and gaps in the alignments, leading to balanced calling of SNPs and INDELs by variant callers. For the experimenter only interested in a specific variant type, it is possible to modify the parameters to provide alignments more sensitive for the variant type of interest. For example, if INDEL discovery is paramount, reducing the Smith-Waterman penalty for the creation and extension of gaps in alignments will lead to a greater likelihood that INDELs will be discovered.

MOSAIKs memory footprint depends on the size of the reference hash-table which, in turn, depends on the hash ($k$-mer) size as well as the length of the reference sequence. For the human genome using the default value of $k=15$, MOSAIK requires approximately 20 GB of memory. For machines with less available RAM, MOSAIK can be run in a low-memory mode that performs alignments chromosome by chromosome. This reduces the required memory to 7 GB, which makes MOSAIK accessible to most machines.

Improvements in the computational performance can be achieved at the expense of decreased sensitivity, but ongoing development (including replacing the traditional Smith-Waterman algorithm with a *single-instruction-multiple-data* (SIMD) Smith-Waterman algorithm (Farrar 2007)) promises significant performance improvements.

# Methods

## Overview

MOSAIK is a hash-based aligner and it hashes reference sequences as its first step. MOSAIK splits the reference sequences into overlapping contiguous *k*-mers (hashes) and stores the positions of each hash in a hash table data structure that guarantees $O(1)$ lookups. Then, MOSAIK hashes each read in the same hash size and looks hashes up in the hash table to obtain the genomic positions of the hashes of a read. Next, nearby hash positions are consolidated as a hash region (hashes of a read may be clustered as several hash regions) where a Smith-Waterman algorithm is applied to align the read to the local region of a genome reference sequence as a final "polishing" step. For paired-end reads, each end-mate of a read is mapped separately. For some cases, that may be one end-mate aligned well and the other one failing to be aligned. The well-aligned mate can be used to try and rescue the unaligned mate using knowledge of the approximate fragment length used in the paired-end read generation.

## Processing Reference Sequences

MOSAIK can handle a nearly unlimited number of reference sequences, however, the maximum aggregated reference length is four billion bases. Alignments to the human transcriptome using more than 95,000 individual reference sequences are easily handled. The available hash sizes are 4-32.

MOSAIK supports the full set of IUPAC ambiguous nucleotide characters. This allows users to use reference sequences that have been masked by confirmed dbSNP (http://www.ncbi.nlm.nih.gov/projects/SNP/) calls. The ambiguity codes minimize the alignment bias that might be caused when aligning to reference sequences containing SNPs. For considering IUPAC, MOSAIK substitutes ambiguous codes with all of the alternative bases represented by the ambiguity code and stores the resulting hashes in the hash table. In order to avoid increasing the size of the jump database dramatically, the ambiguity codes N and X are not considered when hashing the reference sequences.

## Clustering Hashes

MOSAIK supports various read formats (SRF, FASTA, FASTQ, Bustard, and Gerald). In each case, the reads are split into a set of overlapping hashes and the genomic positions of each hash are queried from the stored reference hash table. A modified AVL tree (Adelson-Velskii and Landis 1962) is employed to cluster nearby hash positions to form a hash region. The clustering algorithm considers sequencing errors, SNPs, and single-based INDELs. For example, consider a 35 base read split into hashes of 15 bases. The first hash consists of the first 15 bases in the read. The second hash consists of bases 2-16 in the read and so on. The read consists of 22 individual hashes, each of which is associated with positions within the reference genome. If the read aligns perfectly to the somewhere in the reference genome (i.e. there are no sequencing errors or variations), each of the 22 hashes will have a reference genome position offset by a single base (i.e. if the first hash in the read is associated with the reference position *x*, the second hash with the reference position *x+1* etc.), the AVL tree will consolidate those hits into a single alignment candidate region (see Supplemental Figure S5(A)). The presence of a single sequencing error will ensure that 15 of the hashes (each hash overlapping the error), will not be associated with the correct genomic coordinate. Since the clustering algorithm considers sequencing errors, however, an alignment candidate region is still present in the AVL tree (see Supplemental Figure S5(B)).

## Applying Smith-Waterman Algorithm

After identifying alignment candidate regions, MOSAIK employs a Smith-Waterman algorithm to align the alignment candidate to the reference genome. The Smith-Waterman algorithm, which was invented over 30 years ago, is still regarded as the most accurate pairwise alignment algorithm and the preferred choice for aligning gapped sequences since it seeks all possible frames of alignment with all possible gaps. Specifically, the alignments are performed using the Smith-Waterman-Gotoh alignment algorithm (Smith and Waterman 1981; Gotoh 1982). A known error mode in reads generated using the Roche 454 sequencers is an uncertainty in the length of homopolymer regions. Thus, when aligning Roche 454 reads, a modified Smith-Waterman scoring matrix is used that assigns a lower gap open penalty in homopolymer regions.

The time complexity of the Smith-Waterman algorithm is $O(n^2)$ which, given the large numbers of short reads being produced by next-generation sequencing technologies, may render the aligner useless due to poor performance. To address this, a banded Smith-Waterman algorithm (Chao et al. 1992) has been implemented to improve the performance. According to our experiments, the runtimes for aligning Illumina and Roche 454 data are reduced by approximately 3x and 8x respectively.

## Rescuing Paired-End Mates

Each mate in a paired-end read is initially aligned individually. There are various factors that lead to some reads failing to be aligned to the reference. In the case of paired-end reads, the aligned mate can be used to try and rescue the unaligned mate using knowledge of the approximate fragment length used in the paired-end read generation. A local alignment search algorithm has been implemented which performs a Smith-Waterman algorithm in the region proximal to the aligned mate. If the read exhibits the expected strand, orientation, and fragment length, the read is considered rescued. Even if both mates in the pair are successfully aligned, the local alignment search may still be triggered, if the alignments are inconsistent with the expected fragment length.

The number of mates rescued by the local alignment search depends largely on the read lengths considered. With increasing read length, the aligner is less likely to miss a potential alignment and therefore fewer alignments are rescued.

## Handling AB SOLiD reads

AB SOLiD reads are represented in colorspace rather than in the more conventional basespace. Most downstream applications do not support colorspace and so alignments should be output in basespace for maximum use to the user. MOSAIK is equipped to align colorspace reads against a colorspace reference and then convert the resulting alignments into basespace for output to the BAM file. The di-base quality conversion algorithm uses the minimum of the two qualities that overlap a nucleotide in basespace. This approach allows users to specify parameters, such as the maximum number of mismatches. Additionally, it enables users to merge aligned SOLiD datasets with datasets from other sequencing technologies.

## Known-Sequence Insertion Detections

MOSAIK is aware of user-specified insertion sequences, e.g. mobile element insertions. When the insertion sequences are provided, the reference hashes are prioritized such that alignment to the given insertion sequences are attempted prior to alignment to the genome reference. An additional tag in the BAM file (the ZA tag) then indicates any alignments of a read hitting the given insertion sequences. Since MEIs are repetitive elements, a read from an MEI can be mapped to several locations within the genome (potentially hundreds of locations). The ZA tag then populated with valuable information about the reads mate, including location, mapping quality and number of mapping locations for the mate. This information ensures that multiple BAM search operations (which can be lengthy for large BAM files) can be avoided. The downstream MEI detector can detect MEI by using ZA tag easily.

Lee *et al.*

# Supplemental Methods

**Retraining Mapping Quality Neural Network**
We simulated 6 million paired-end reads from *E. coli* genome and aligned them by MOSAIK with "-zn" option that attach ZN optional tag for each alignment in the output bam-formatted file. A ZN tag consists of the best Smith-Waterman score, the next-best Smith-Waterman score, entropy of the read, length of the longest perfect match, the number of obtained alignments, and the number of obtained hashes. We further attached XC optional tag that shows the correct positions of reads. Then, we applied our training program to the files that have been attached ZN and XC. An example is given on
https://github.com/wanpinglee/MOSAIK/blob/master/demo/RetrainMQ.sh.

**Detecting Specified Insertion Sequences**
MOSAIK can be aware of alignments mapped to the specified insertion sequences. In a reference fasta-formatted file, users should attach insertion sequences with the same prefix. For example, we attached the mobile element insertions (MEIs) with the prefix "moblist_" after human genome references. Then, when aligning reads, we enable "-sref moblist" option to let MOSAIK move hashes located in the MEIs to the top. Consequently, aligning reads to MEIs is a priority by MOSAIK. The other option "-srefn <int>" can limit the number of hashes moved to the top. Once alignments are mapped to the specified insertions, MOSAIK will indicate that in ZA optional tags.

# Supplemental Figures

**Figure S1:** The distributions of alignments' softclips.

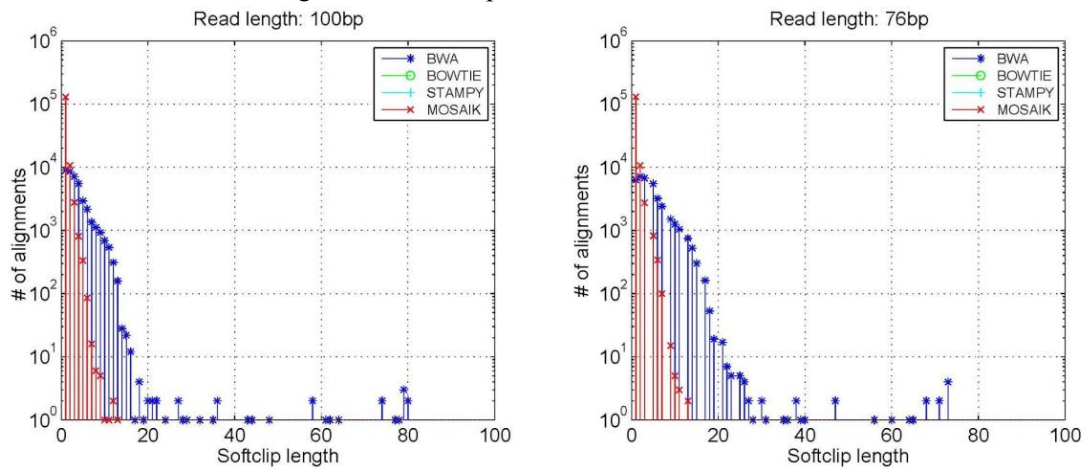

**Figure S2:** The complete information of Figure 1. The positive predictive value of aligners (the number of correctly mapped reads divided by the total number of mapped reads) as a function of mapping quality threshold. Datasets in (A) 100bp and (B) 76bp read lengths. PPV, TP, and FP stand for positive predictive value, true positive, and false positive, respectively.

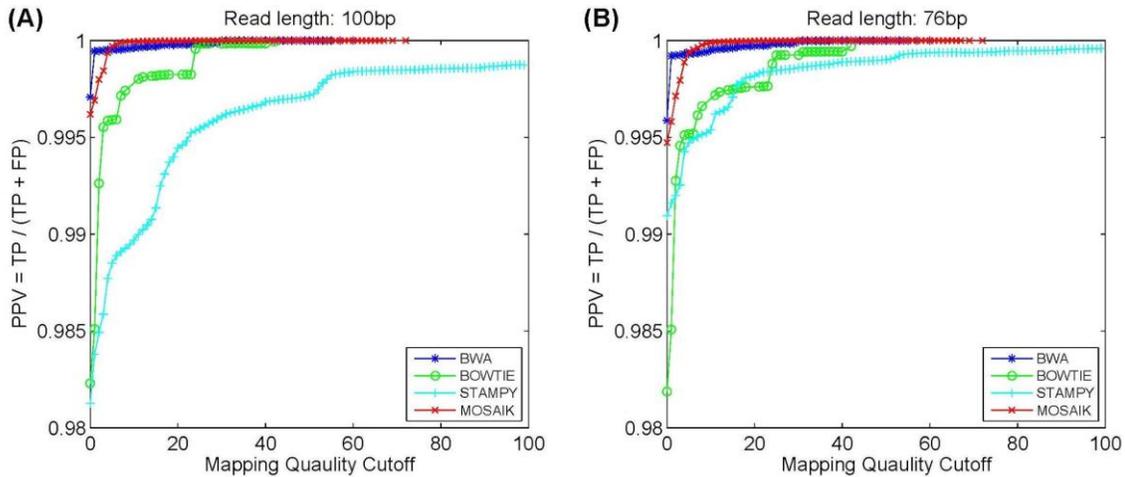

**Figure S3:** The complete information of Figure 5. The receiver operating characteristic (ROC) curves of SNPs called by FREEBAYES and SAMTOOLS. The points on the curves are sorted by called qualities and the points closer to the upper-right corner have higher called qualities. The true positive (TP), false positive (FP), and false negative (FN) are calculated by intersecting SNPs called on each aligner's alignments and gold SNPs called on the simulated alignments.

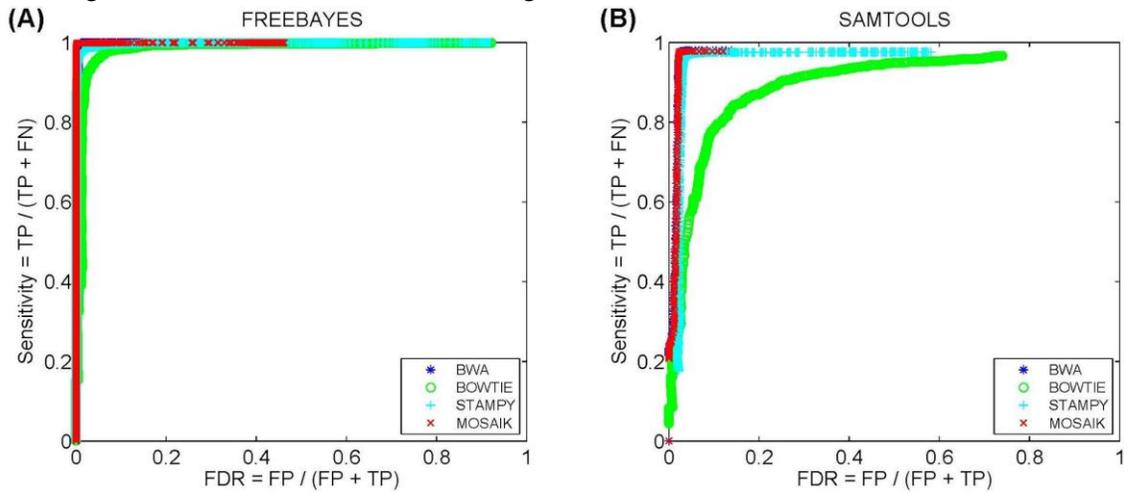

**Figure S4:** The short INDELs that are inserted for investigating the aligners' abilities for them, and the read coverage for each length INDEL.

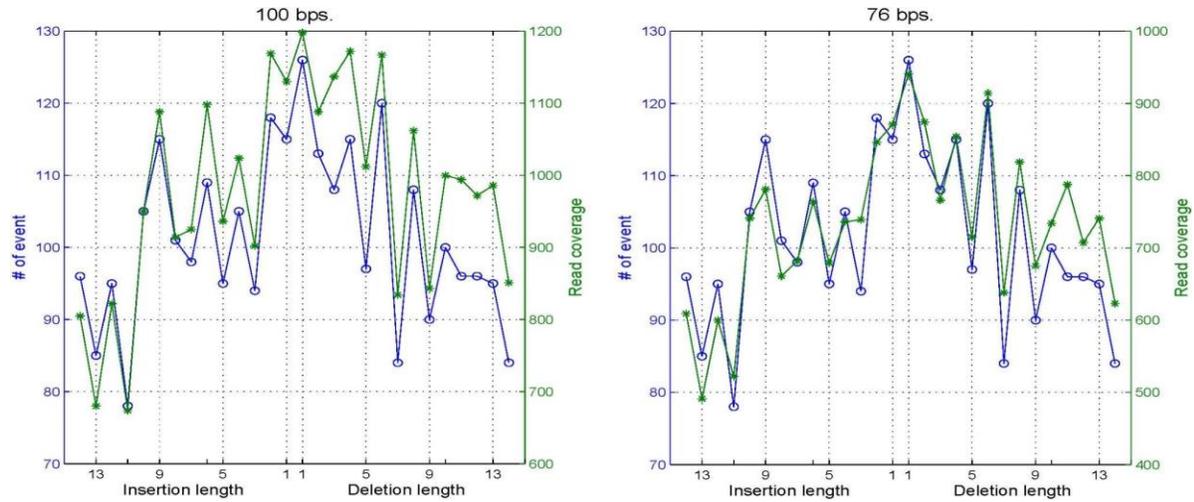

**Figure S5:** MOSAIK hash clustering. (A) The read uniquely aligns perfectly to the references, all hashes will succeed in finding the adjacent reference locations and the AVL tree will consolidate those hashes into one alignment candidate region. (B) However, if only one hash succeeds in finding the proper reference location because of sequencing errors, an alignment candidate region is still present in the AVL tree.

**(A)**

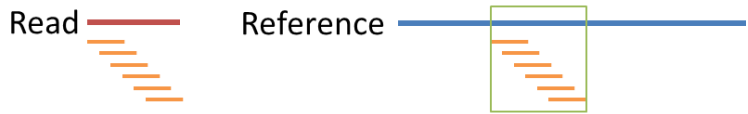

**(B)**

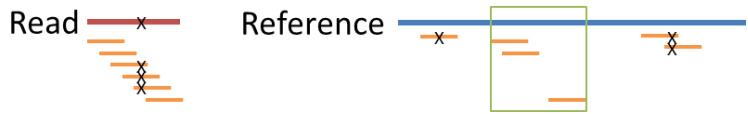